\documentclass[11pt]{article}
\usepackage{graphicx}
\usepackage{url}

\author{
Nadine Kroher \\
\texttt{Department of Applied Mathematics II, University of Seville}\\
\and 
Aggelos Pikrakis\\
\texttt{Department of Informatics, University of Piraeus}\\
}
\title{Exploratory Analysis of a Large Flamenco Corpus using an Ensemble of Convolutional Neural Networks as a Structural Annotation Backend} 

\begin{document}
\maketitle
\begin{abstract}
We present computational tools that we developed for the analysis of a large corpus of flamenco music recordings, along with the related exploratory findings. The proposed computational backend is based on a set of Convolutional Neural Networks that provide the structural annotation of each music recording with respect to the presence of vocals, guitar and hand-clapping (``palmas''). The resulting, automatically extracted  annotations, allowed for the visualization of music recordings in structurally meaningful ways, the extraction of global statistics related to the instrumentation of flamenco music, the detection of a cappella and instrumental recordings for which no such information existed, the investigation of differences in structure and instrumentation across styles and the study of tonality across instrumentation and styles. The reported findings show that it is feasible to perform a large scale analysis of flamenco music with state-of-the-art classification technology and produce automatically extracted descriptors that are both musicologically valid and useful, in the sense that they can enhance conventional metadata schemes and assist bridging the semantic gap between audio recordings and high-level musicological concepts.
\end{abstract}

\section{Introduction}\label{sec:intro}
The growing availability of digital music collections, in the form of symbolic music representations and audio recordings, along with recent advances in symbolic and audio-based Music Information Retrieval (MIR), have enabled researchers to scale descriptive music analysis \cite{descriptive1,descriptive2} from small collections of a few manually annotated items to large corpora of music content. In the case of  non-Western music traditions, which often lack formal documentation, exploratory studies of music corpora are a powerful means to reveal distinctive musical characteristics, validate musicological research hypotheses and identify meaningful features for the automatic organization of digital collections. It is now acknowledged that the development of computational analysis tools  can contribute significantly to the preservation of cultural heritage, assist the understanding of the past and help bridge the semantic gap between sparsely documented content and concepts at various levels of music abstraction. %citation here

Data-driven studies have already shown significant potential in discovering knowledge and hidden correlations within and among music genres. The work in \cite{worldMusicSinging}, which was based on a learned dictionary of vocal pitch contour elements, showed that singing styles with geographical proximity, exhibit similar characteristics. In \cite{distinctivePatterns}, a comparison of melodic pattern occurrences between genre-specific corpora and an anti-corpus, revealed distinctive patterns for different folk song traditions. Using Subgroup Discovery techniques, the method in \cite{subgroups} generated a set of interpretable rules which can associate folk song transcriptions with their origin. An analysis of manually annotated harmonic progressions in rock music \cite{rockHarmony} revealed a number of frequently occurring chords and chord sequences, as well as evolutionary trends over several decades. In \cite{ragtime}, a large-scale corpus study of ragtime scores led to the empirical validation of certain hypotheses that are related to the  occurrence of rhythmic patterns. 

In this paper, we present, for the first time, an exploratory study of a large flamenco corpus \cite{corpusCOFLA}, from a computational analysis perspective. In particular, we focus on the musical structure with respect to instrumentation and its relation to  style and tonality. Flamenco is a rich oral music tradition from Southern Spain, which, under the influence of diverse cultures, has evolved from its folkloric origin to an elaborate art form \cite{bridges}. As an oral tradition with a highly improvisational character, flamenco is not bound to scores, but it is based on style-specific, melodic, rhythmic and harmonic templates, which set the basis for expressive performance. These underlying rules are, to a large extent, implicitly passed from generation to generation, and, despite recent efforts to formalize flamenco music theory \cite{flamenco,fernandez}, many aspects remain undocumented.   

Given the absence of scores, our large-scale analysis  of flamenco music relies on automatic annotations that are directly extracted from audio recordings. More specifically, we developed an automatic annotation system which segments each music recording into sections that are consistent with respect to instrumentation. This is feasible, because in classical flamenco recordings, instrumentation is in most cases limited to singing with guitar accompaniment and rhythmic hand-clapping (referred to as \textit{palmas}). Some styles are traditionally performed a capella, and, in rare cases, performances can be instrumental. 

To detect segments with consistent instrumentation, we employ an ensemble of Convolutional Neural Networks (CNNs), where each CNN solves a binary classification problem. CNNs were selected as a backend, because, over the past few years, they have been successfully applied to various analysis tasks, ranging from face detection \cite{faceDetection,faceDetection2} and object recognition \cite{imagenet}, to MIR tasks including boundary detection \cite{boundary}, music recommendation \cite{recommender} and instrument recognition \cite{instrRec}.

\begin{figure}[ht]
\centering
     \includegraphics[width=0.8\textwidth]{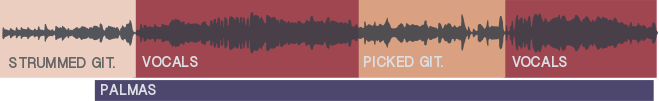}
      \caption{Schematic illustration of the segmentation of a flamenco recording with respect to instrumentation.}
       \label{fig:teaser}
\end{figure}

The three CNNs proceed to the segmentation of an audio recording as follows: 
\begin{itemize}
\item The first classifier detects parts of the recording where the singing voice is present. 
\item In the remaining solo guitar sections, a second classifier distinguishes between two guitar playing techniques, \textit{strumming} and \textit{picking}. 
\item The third classifier detects parts of the recording where the \textit{palmas} are present, irrespective of the outcome of the other classifiers. 
\end{itemize}
Figure \ref{fig:teaser} depicts schematically the outcome of the annotation procedure.
This type of automatic annotation sets the basis for a large-scale analysis of a music corpus containing more than $1500$ commercial flamenco recordings \cite{corpusCOFLA}. In particular: 
\begin{itemize}
\item We detect instrumental and a cappella recordings based on the estimated percentage of  vocal and non-vocal frames. 
\item We explore intra and inter-style similarity based on two types of features: song-level instrumentation statistics and a time series representation of the temporal structure. We also investigate the usefulness of these features for automatic style recognition tasks.
\item We verify music theoretical assumptions on tonality with respect to style and instrumentation, by correlating pitch class profiles with mode templates.
\end{itemize}

The results of this study can be further exploited by several potential applications for automatic indexing and content-based retrieval in digital flamenco collections, can provide valuable clues towards the unresolved problem of automatic style detection \cite{styleDetection}, and can reveal interesting musicological observations.

The rest of the paper is structured as follows: Section \ref{sec:backend} provides a detailed description and evaluation of the structural segmentation backend. The  exploratory analysis study, its results and a detailed description of the corpus, are presented in Section \ref{sec:mining}. Finally, conclusions are drawn in Section \ref{sec:conclusions}.

\section{Structural annotation backend}
\label{sec:backend}
\begin{figure}[ht]
\centering
     \includegraphics[width=0.8\textwidth]{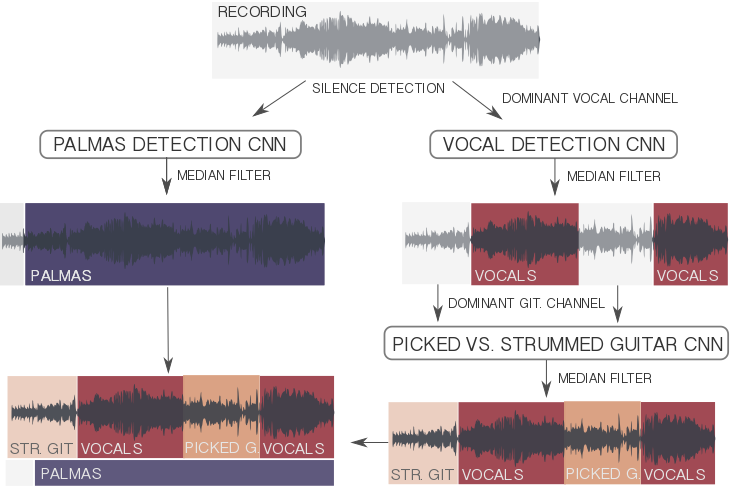}
      \caption{Schematic overview of the segmentation backend.}
       \label{fig:backend}
\end{figure}

In this section, we describe the segmentation backend, the purpose of which is to break a flamenco music recording into  segments of consistent instrumentation, thus yielding a structural annotation of the track. An overview of the proposed multi-stage system architecture is shown in Figure \ref{fig:backend}. 

At a first stage, given a raw audio recording,  silent sections are detected based on the short-term energy envelop of the signal. Silence usually occurs at the endpoints of a recording and between successive sung sections in a cappella styles. The detected silent parts are excluded from  subsequent classification stages to improve classification accuracy.

Subsequently, we exploit the fact that in a flamenco stereo recording, the vocals are usually dominant in one channel and the guitar in the other. It is worth noting that this is not only the case with live recordings, where the resulting stereo panorama corresponds to the physical location of singer and guitarist on stage, but it is also common practice in multi-track flamenco studio productions, as a means to separate voice and guitar in the artificially created panorama. Here, we apply the heuristic that was proposed in \cite{cante} to identify the channel in which the vocals are dominant. In the sequel, we will refer to this channel as the \textit{vocal channel} and to the other one as the \textit{guitar channel}. Furthermore, we create a mix of the two channels by computing the mean signal, to which we will later refer as the \textit{mono mix}. If the recording is  monophonic, the channel selection stage is skipped, in which case, the vocal and guitar channels coincide with the mono mix signal. 

At the next stage, the vocal channel is passed as input to the first CNN, which acts as a vocal detector and estimates, on a frame-level basis, if the singing voice is present or not. Successive frames that have been classified to the vocal class are considered to form a vocal segment and are annotated accordingly. Vocal detection in flamenco music has been previously studied in the context of automatic transcription \cite{cante} and in an unsupervised scenario \cite{unsupervised}. However, the CNN-based method that is proposed in this paper is shown to outperform previous approaches. The remaining, non-vocal segments are taken from the guitar channel and are passed as input to the next CNN, which discriminates, again on a frame-level basis, between two guitar playing techniques, namely, \textit{picking} and \textit{strumming}. The third CNN operates independently, takes the mono mix signal as input and detects, for each frame, if \textit{palmas} are present or not. 

Below, we describe in detail, from a technical perspective, each stage of the processing chain and justify the respective design choices.

\subsection{Pre-processing}

After normalizing the audio recording and re-sampling it to $44100 \mathrm{Hz}$, we detect silent frames by analyzing the signal energy over non-overlapping short-term windows, $2048$ samples long. A frame is considered to be silent, if its energy value is at least $35$ $\mathrm{dB}$ lower than the  highest energy value in the recording. 

In order to identify the channel in which the vocals are dominant, we apply the heuristic that was proposed in \cite{cante}, which evolves around the observation that the presence of the singing voice increases the energy in the frequency range between $500$  $\mathrm{Hz}$ and $6$  $\mathrm{kHz}$. Although this criterion is not sufficient to accurately detect the presence of vocals at the frame-level, it has however shown to reliably point to the channel in which the vocals are dominant. For each channel, we compute, for every  short-term frame, the ratio of the sum of  magnitudes of the DFT coefficients in the frequency range from $500$  $\mathrm{Hz}$ to $6$ $\mathrm{kHz}$, to the respective sum in a lower frequency band, ranging from $80$  $\mathrm{Hz}$ to $400$  $\mathrm{Hz}$. The channel with the higher average ratio is selected as the vocal channel and the other one as the guitar channel.  

\subsection{Feature extraction}
Before a signal is fed to a CNN, a short-term feature extraction stage extracts a sequence of feature vectors from its time-domain representation. Specifically, the signal is first parsed with a non-overlapping moving window, $2048$ samples long ($46.4$ ms). At each window, the Discrete Fourier Transform (DFT) is computed and it is given as input to a mel filter-bank that yields $128$ mel-band energy coefficients. Each mel-filter computes a weighted sum of the DFT coefficients that lie inside its frequency range (\cite{mel}).

\begin{figure}[ht]
\centering
     \includegraphics[width=0.6\textwidth]{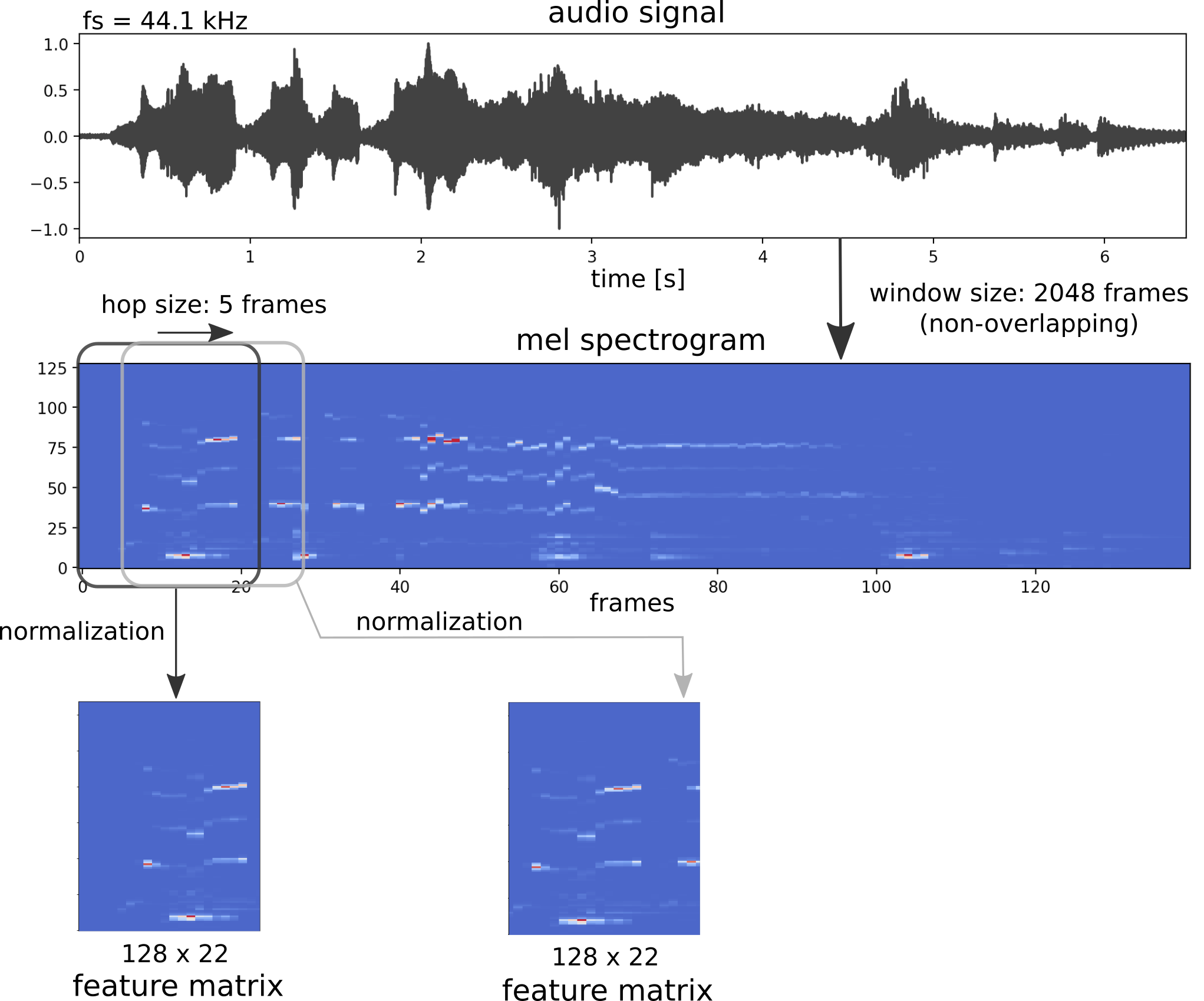}
      \caption{Illustration of the feature extraction stage.}
       \label{fig:featureExtraction}
\end{figure}

After the short-term feature extraction stage has been completed, a subsequence of $22$ successive feature vectors is aggregated each time to form  a $128\times22$ image representation, that will serve to assign a classification label to the time instant corresponding to the middle of the subsequence. As a result, the short-term feature sequence generates a sequence of 2-D images. Each image reflects the evolution of spectral content (image height) over a neighborhood of frames (image width) around the time instant for which a classification decision will be made. Given the short-term window step of $2048$ samples and the sampling frequency of $44100 \mathrm{Hz}$, an image width of $22$  frames corresponds to $\approx 1 s$ of audio, i.e., to $\approx 0.5$ s of audio around the time instant to which the classification label will be assigned. In order to reduce the computational burden, the image slides by a hop size of $5$ frames at a time. Equivalently, a classification decision is taken approximately every $0.25 $ s. To ensure robustness to signal intensity variation, each image is normalized by dividing each one of its elements by the maximum absolute value of the image. The feature extraction process is depicted in Figure \ref{fig:featureExtraction}.

This approach is inspired by the two-dimensional pixel matrices that are used in image classification systems and has shown to give promising results in various audio analysis tasks (see for example \cite{boundary,instrRec,soundtracks}). 
For the CNN that serves as the \textit{palmas detector}, the logarithm of the mel-band energy values is computed, to capture the characteristic sudden energy increases that are produced by the percussive impulses. % do we normalize here? 

\subsection{CNN architecture}
\begin{figure}[ht]
\centering
     \includegraphics[width=0.8\textwidth]{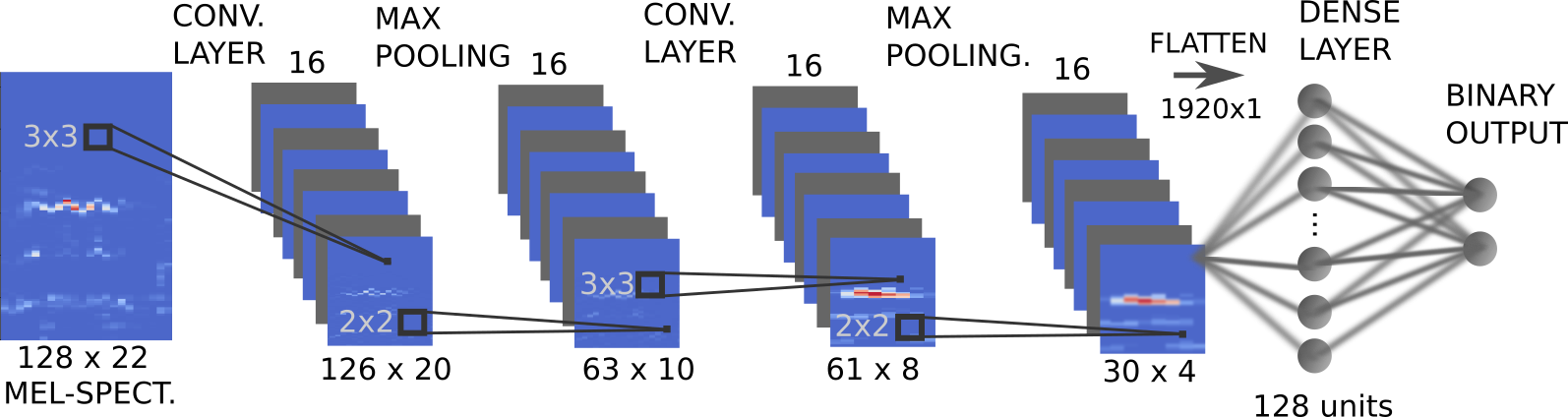}
      \caption{Illustration of the CNN architecture.}
       \label{fig:arch}
\end{figure}

The three CNNs follow a common architecture, shown in Figure \ref{fig:arch}. Specifically, each $128\times22$ input matrix, for which a classification decision will be taken, is passed through two convolutional layers. Each layer contains $16$ convolutional masks (feature detectors) of size $3\times3$. The output of each convolutional operation is passed through a \textit{relu} function and the resulting feature matrix is subsequently subsampled by a $2\times2$ max pooling scheme. The resulting $16\times30\times4$ representation at the output of the second layer is unfolded (flattened), yielding a $1920\times1$ one-dimensional representation, which is subsequently fed as input to a standard, fully connected feed forward neural network. This network has a hidden  layer of $128$ units and a softmax output layer of two units due to the binary nature of the classification tasks under study. Figure \ref{fig:arch}  presents the data flow across the adopted neural network architecture, including  the dimensionality of matrices at each processing stage.

\subsection{Post-processing}
As it was previously described, when an input image is given, the predicted class is interpreted as the label of the time instant corresponding to the middle of the respective frame subsequence. Furthermore, assuming that instrumentation remains consistent over short time periods, we apply a median-filter to the sequence of binary classification decisions to suppress noisiness. In particular, for the vocal detection and guitar playing technique classification tasks, a median filter, $1 \mathrm{s}$ long, is used. For the \textit{palmas} detection module, we use a longer median filter, $5 \mathrm{s}$ long, as we expect that the  \textit{palmas} events unfold over longer time periods.

\subsection{Classifier training}
The singing voice classifier is trained and evaluated on the \textit{cante100} dataset, which has been previously used in the context of vocal detection in flamenco music \cite{unsupervised}. The collection is a subset of 100 commercial recordings taken from the \textit{corpusCOFLA} collection of flamenco recordings \cite{corpusCOFLA}. For the classifier of guitar playing techniques, we manually annotated strummed and picked guitar sections on the same dataset. It has to be noted that, with respect to the presence of \textit{palmas}, the \textit{cante100} collection proves to be rather unbalanced - only about 10 \% of the tracks contain hand-clapping. We therefore assembled a  balanced subset of $100$ recordings from the \textit{corpusCOFLA} collection, where half the recordings contain hand clapping. 

All three classifiers are trained and evaluated using a $10$-fold cross-validation scheme. For the singing voice detection task, artist-filtered folds are used, i.e., all songs of the same singer are included in the same fold. For the other two tasks, folds are song-filtered, ensuring that all training images from a common recording are located in the same fold. These standard measures are taken to reduce the risk that the classifiers will overfit to the timbral characteristics of a particular singer or recording. After the feature extraction stage is carried out on the aforementioned datasets, each classifier is trained for $50$ epochs using the \textit{Adam} \cite{adam} algorithm for optimization of the cross-entropy function over the training set. During each training epoch, the images are shuffled and are grouped to form mini-batches ($128$ images per mini-batch). An early-stopping criterion is also used, which terminates the training procedure if the loss-value does not decrease  significantly during a patience-period of $5$ epochs (at least by a value of $0.01$).

\subsection{Baseline methods}
The performance of each binary classifier is compared against at least one baseline system. More specifically, the vocal detection classifier is compared against: 
\begin{itemize}
\item{The method in \cite{gmmBL} which is based on a Gaussian Mixture Model (GMM) per class, trained on frame-level mel-frequency cepstral coefficients \textit{(GMM-BL-VOC)}}.
\item{The vocal detection module of the automatic transcription system \textit{cante} \cite{cante}, where vocal sections are first detected depending on the presence of the predominant melody and are subsequently filtered based on low-level audio descriptors \textit{(CANTE)}}.
\item{The unsupervised method in \cite{unsupervised}, which learns a dictionary of atoms from the bark band representation of the signal and detects the singing voice based on the activation patterns of the computed dictionary atoms \textit{(UNSUPERVISED)}}.
\end{itemize}

As a baseline method for the performance of the guitar playing technique and \textit{palmas} detection classifiers, we implemented a GMM-based method for each task. Specifically, we extract the mel-frequency cepstrum coefficients and their first and second derivatives over time in $23$ $\mathrm{ms}$ long windows, with a hop size of $10$ $\mathrm{ms}$ and average the extracted short-term sequence with a $1$ s long moving window. We then train a GMM with $16$ components for each class. Each frame is assigned to the class whose GMM yields the highest probability.   

All baseline methods and the CNN approach are trained and evaluated using the same $10$-fold cross validation scheme. Furthermore, the same pre- and post-processing stages are applied as for the proposed system.

\subsection{Experimental results}
The results of the vocal detection experiments are shown in Table \ref{tab:vocDetResults}. It can be seen that the proposed CNN-based method achieves the highest F-measure ($0.95$) among the four algorithms. A manual analysis showed that errors were mainly caused by shouts from the guitarist and audience during performance, which were mistakenly classified as singing voice segments. Although limited in duration, this is a typical phenomenon in flamenco performances.

Tables \ref{tab:falDetResults} and \ref{tab:palDetResults} show the results of the guitar playing technique and \textit{palmas} classification experiments. For both tasks we observe that the proposed method outperforms the GMM-based baseline approach. For the guitar playing technique classification, the CNN yields an F-measure of $0.91$ and for the \textit{palmas} detection task an F-measure of $0.95$. The performance of the GMM-baseline method is significantly lower, yielding an f-measure of $0.74$ for the guitar playing technique classification task and $0.66$ for the \textit{palmas} detection method. In the former, \textit{picked} guitar was considered as the positive class. 

We furthermore observe, that the performance of each CNN-based classifier is balanced with respect to precision and recall, indicating that there is no notable bias towards one of the target classes. In the vocal detection task, the \textit{CANTE} method reaches the highest recall value of $0.97$, however, at the cost of a significantly lower recall value of $0.75$, compared to the proposed method ($0.95$). 

\begin{table}
\caption{Experimental results: vocal detection.}
\label{tab:vocDetResults}
\begin{center}
\begin{tabular}{lccc}
  \hline
  Method & Precision & Recall & F-measure\\
  \hline
  \textit{GMM-BL-VOC} & 0.92 & 0.85 & 0.88\\
  \textit{CANTE} & \textbf{0.97} & 0.75 & 0.85\\
  \textit{UNSUPERVISED} & 0.94 & 0.77 & 0.85\\
  \textit{PROPOSED} & 0.94 & \textbf{0.95} & \textbf{0.95}\\
\hline
\end{tabular}
\end{center}
\end{table}

\begin{table}
        \caption{Experimental results: \textit{palmas} detection.}
\centering
  \begin{tabular}[t]{lccc}    
  		\hline
  		Method & Precision & Recall & F-measure\\
  		\hline
  		\textit{GMM-BL-PAL} & 0.66 & 0.65 & 0.66\\
  		\textit{PROPOSED} & \textbf{0.94} & \textbf{0.95} & \textbf{0.95}\\
		\hline
		\end{tabular}
        \label{tab:palDetResults}
\end{table}

  \begin{table}
    \caption{Experimental results: guitar playing technique classification.}
  \centering  
        \begin{tabular}[t]{lccc}
  		\hline
  		Method & Precision & Recall & F-measure\\
  		\hline
  		\textit{GMM-BL-GIT} & 0.69 & 0.80 & 0.74\\
  		\textit{PROPOSED} & \textbf{0.91} & \textbf{0.92} & \textbf{0.91}\\
		\hline
		\end{tabular}
        \label{tab:falDetResults}
\end{table}

\subsection{What did the networks learn?}
\begin{figure}[!htbp]
\centering
     \includegraphics[width=0.75\textwidth]{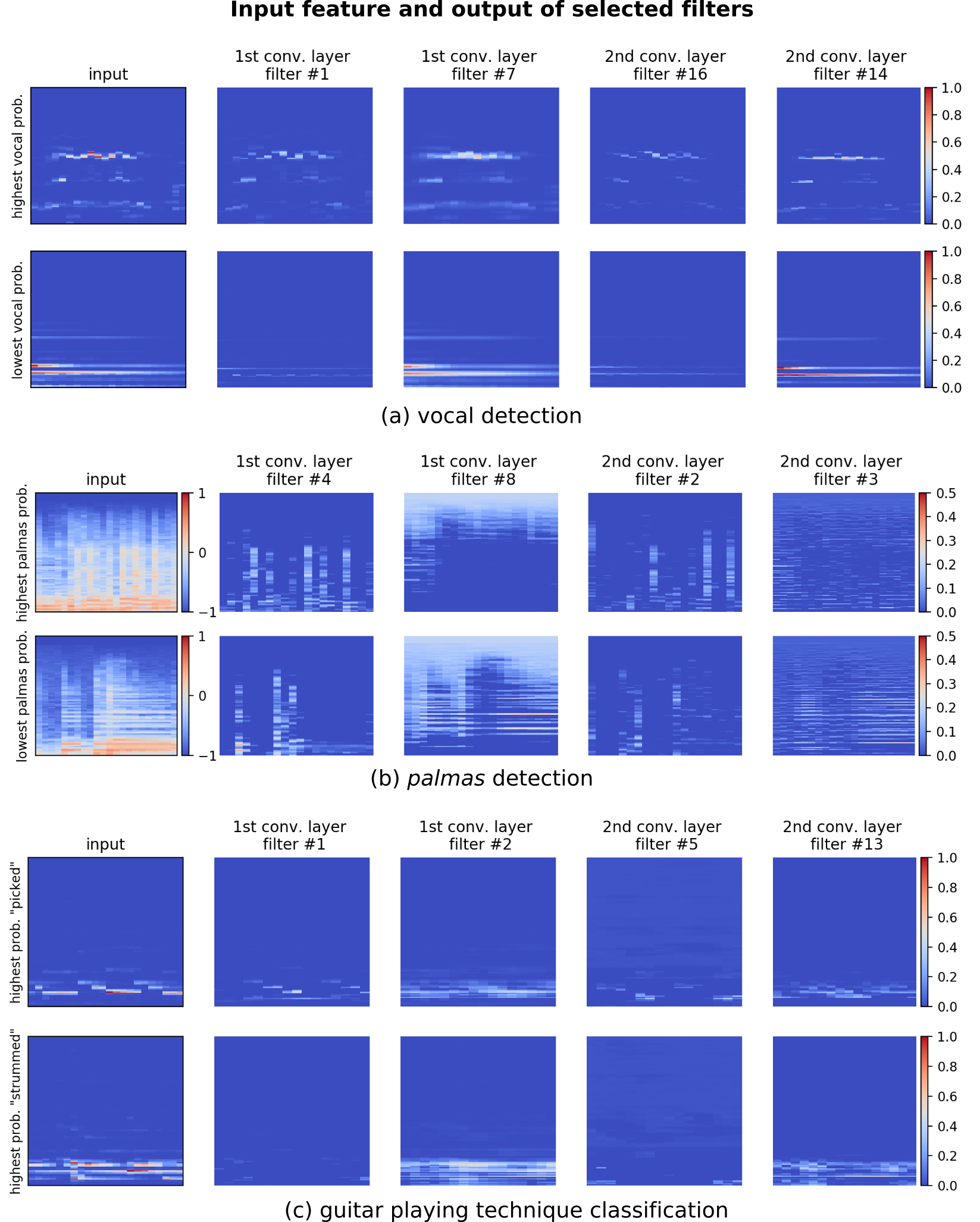}
      \caption{Visualisation of input features and first and second layer convolutional filter outputs: (a) vocal detection; (b) \textit{palmas} detection; (c) guitar playing technique classification.}
       \label{fig:selectedFilters}
\end{figure}

\begin{figure}[!htbp]
\centering
     \includegraphics[width=0.75\textwidth]{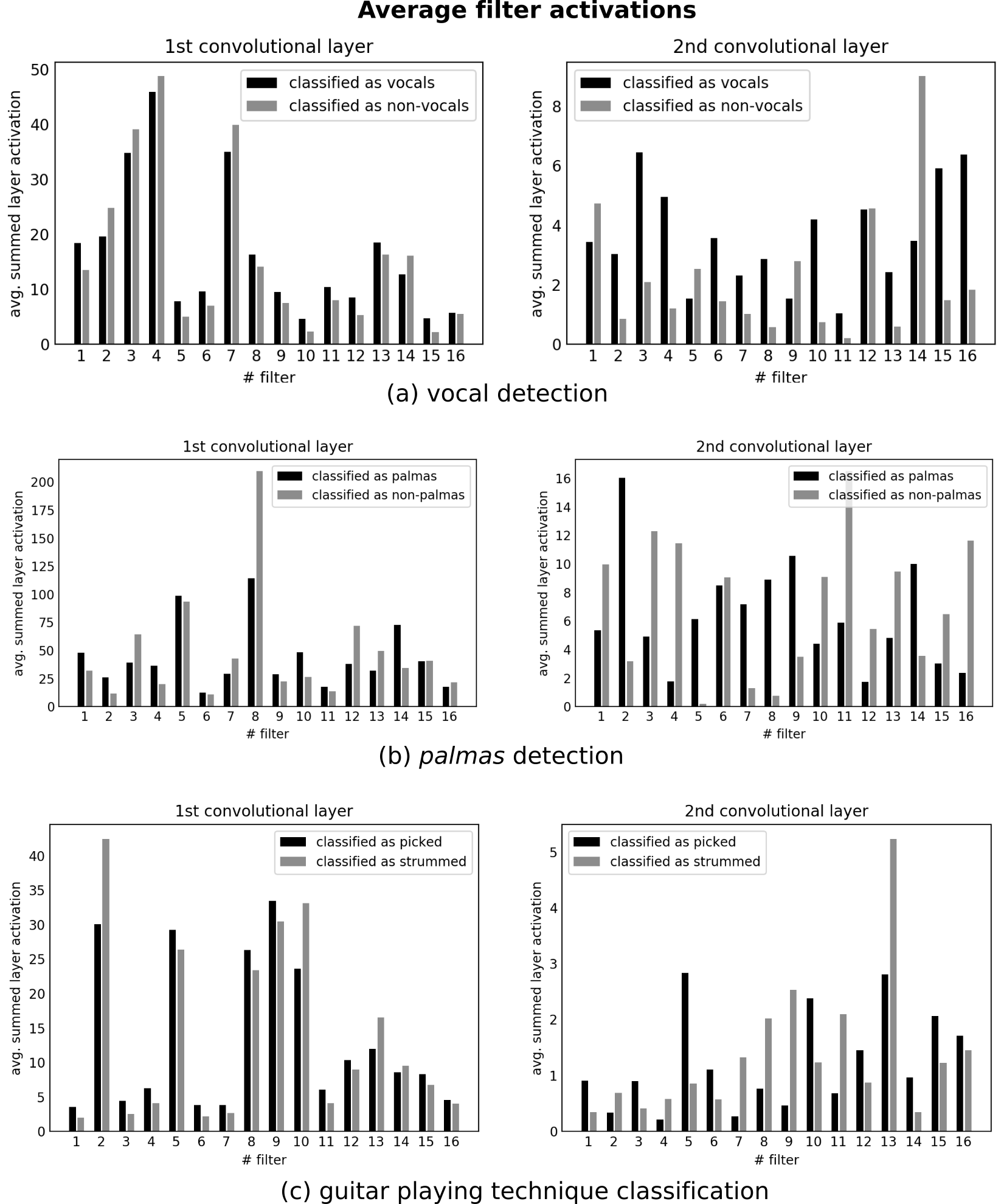}
      \caption{Average activation of first and second layer convolutional filters: (a) vocal detection; (b) \textit{palmas} detection; (c) guitar playing technique classification.}
       \label{fig:filterActivations}
\end{figure}

We are now making an attempt to analyze in detail how the networks solve the tasks at hand, which is always a challenging issue when convolutional architectures are employed. To this end, based on selected recordings, we examine the \textit{relu} activations of the convolutional filter output of the first and second layer (Figure \ref{fig:filterActivations}) for each target class. Furthermore, we visualize the output of selected filters from both layers for the input frames which yield the highest and lowest output probabilities for each target class (Figure \ref{fig:selectedFilters}). 

To begin with, an examination of the two input images that yield the highest and lowest posterior probability for the singing voice class (Figure \ref{fig:selectedFilters}), reveals that the presence of the singing voice causes the spectral energy to be centered in a higher frequency range, compared to the case of  guitar. 
We can furthermore observe a fluctuation of the dominant frequency bins over time, which is not the case with the horizontal spectral lines of the guitar image (which do not exhibit fluctuation). Comparing images with and without \textit{palmas}, we can see, that hand clapping produces sudden short-duration increases in the spectral magnitude, that become more apparent in high frequencies. Filter no. 4 seems to detect these sudden bursts because of the vertical stripes visible in the output its mask. Filter no. 8 yields a high activation in areas with low spectral magnitude. Consequently, this filter has a high activation for frames without \textit{palmas}, that exhibit less energy in the high frequency range. The feature matrices representative for strummed and picked guitar differ mainly in the number of temporally simultaneous peaks and the long-term pitch fluctuation. The strummed guitar produces multiple sustained peaks, visible as horizontal lines, whereas the picked guitar results in single peaks which quickly decay. 

Figure \ref{fig:filterActivations} (a) shows the average activation of the convolutional filters of the first and second layer, for images which are classified as containing / not containing vocals. 

It can be seen that each filter tends to have a higher activation for one of the two classes. This difference in average activation is more apparent in the second layer. A similar behavior can be observed for the \textit{palmas} detection (Figure \ref{fig:filterActivations} (b)) and guitar playing style classification (Figure \ref{fig:filterActivations} (c)) tasks. This observation is furthermore reflected in the output of selected filters for frames representative of the output classes (Figure \ref{fig:selectedFilters}). Consequently, each filter seems to specialize in one of the two output classes.

\section{Exploratory corpus study}\label{sec:mining}
In this section, we employ the previously described structural segmentation backend to explore a large corpus of flamenco music recordings. More specifically, we demonstrate how the automatically extracted annotations can be used to (a) augment editorial meta-data with meaningful content-based descriptors and visualizations, (b) discover genre-specific characteristics and evaluate their potential for music information retrieval tasks, and (c) illustrate music theoretical concepts in a large-scale data-driven context. Our goal is to focus on computationally feasible analysis experiments and tasks that would require immense effort when conducted manually.

First, we show that the visualization of the automatically extracted structure of a recording yields a concise overview of its content and can  reveal interesting musical properties. Then, we compute global statistics over the corpus with respect to instrumentation and  describe how the backend can be used to automatically identify a cappella and instrumental recordings. At a next stage, we examine  how styles differ with respect to instrumentation and respective structure and investigate the potential use of the segmentation backend in automatic style classification systems. In addition, we illustrate the relationship between style, instrumentation and tonality via an example of three popular styles with distinct harmonic characteristics.

\subsection{Corpus}
We focus on Version 1.1 of \textit{corpusCOFLA}\footnote{\url{http://www.cofla-project.com/?page_id=170}} \cite{corpusCOFLA}, a large collection of commercial recordings. This dataset was recently assembled in order to create a representative set of classical flamenco recordings that will enable the realization of  computational analysis tasks. It encompasses a number of renown anthologies, together with their editorial meta-data. The annotated styles, which exhibit varying levels of detail and sub-style distinction, were classified into $86$ categories and artist names were assigned unique identifiers. An overview of the statistics of the corpus is given in Table \ref{tab:corpusStats} and the ten most frequently occurring styles are shown in Table \ref{tab:top10styles}.

\begin{table}
\caption{Statistics of the \textit{corpusCOFLA}.}     
\centering
        \begin{tabular}{ |clc|cl }
           \hline
  			no. recordings  & 1594\\
  			no. styles     & 86\\
  			tracks without style information & 109\\
  			tracks containing multiple styles & 104\\
  			no. singers & 364\\
  			no. anthologies & 10\\  
			\hline
        \end{tabular}
        \label{tab:corpusStats}
 \end{table}

\begin{table}
\caption{10 most frequently occurring styles in the \textit{corpusCOFLA}.}        
\centering
        \begin{tabular}{ |c|clc|c|}
           \hline
\textbf{style} & \textbf{no. of recordings}\\
	\hline
	Fandangos  &   203\\
	Soleares   &   150\\
	Siguiriyas  &  124\\
	Buler{\'i}as    &  120\\
	Malague{\~n}as  &   73\\
	Tangos      &   59\\
	Alegr{\'i}as    &   43\\
	Grana{\'i}nas   &   39\\
	Tarantas    &   37\\
	Tientos     &   34\\
\hline
        \end{tabular}
        \label{tab:top10styles}
\end{table}

\subsection{Visualization of the structural annotations}
Before proceeding with the analysis of the corpus as a whole, we show on the basis of two examples, how the output of the classifiers can provide a compact, content-based visualization of a flamenco recording. 

\begin{figure}[ht]
\centering
     \includegraphics[width=0.99\textwidth]{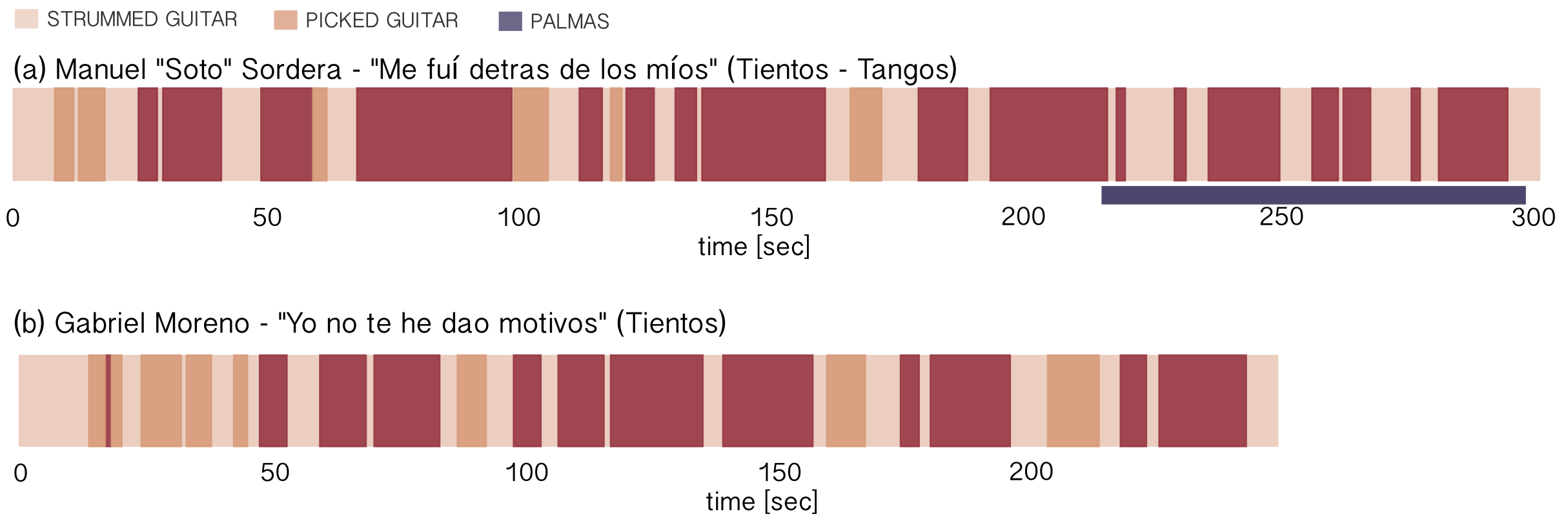}
      \caption{Example of visualizing structure: (a) \textit{tientos - tangos} and (b) \textit{tientos}.}
       \label{fig:vis1}
\end{figure}

Figure \ref{fig:vis1} (a) presents the automatic segmentation of the song \textit{Me fu{\'i} detras de los m{\'i}os}, performed by \textit{Manuel Sordera}. This particular recording, which contains two styles, starts with the \textit{tientos} style and ends with a \textit{tango}. It is not uncommon that these two styles are encountered in a single performance, in this particular order. As a matter of fact, they are closely related styles because they follow a common harmonic progression and rhythmic structure. However, a \textit{tango} is performed at a faster tempo with stronger rhythmic accentuation. 

The visualization illustrates some further interesting differences between these two styles in this  recording. The \textit{tientos} part, which lasts from the beginning until approximately second $220$, is characterized by long vocal segments, with guitar interludes alternating between picked and strummed sections. With the onset of the \textit{tangos} part, the \textit{palmas} set in to emphasize the underlying rhythmic pattern and the vocal sections become shorter. The guitar accompaniment is limited to strumming, which, due to its percussive nature, further increases the accentuation of the rhythm. Figure \ref{fig:vis1} (b) shows the same form of representation for a recording containing only \textit{tientos}. Since in this recording, there is no modulation to \textit{tangos}, there are no \textit{palmas} and the guitar moves between picked and strummed sections until the end of the recording.

\begin{figure}[ht]
\centering
     \includegraphics[width=0.95\textwidth]{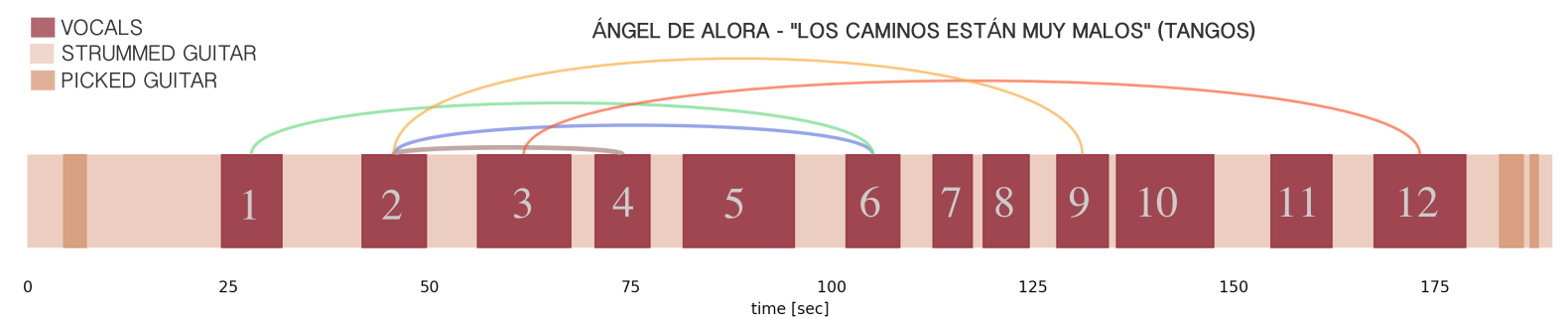}
      \caption{Example of visualizing structure and vocal section similarity.}
       \label{fig:vis2}
\end{figure}

The proposed visualization can  be combined with other content-based descriptors that cover different musical aspects. For example, in Figure \ref{fig:vis2}, structural annotations are combined with descriptors related to melodic repetition. Specifically, the arcs connecting vocal segments correspond to the five highest similarity values among all vocal sections. The thickness of the arc is proportional to the respective similarity value, which is computed using the alignment method  in \cite{audioPatterns}. It can be seen that vocal sections $1$, $2$, $4$, $6$ and $9$ are of similar melodic content and that section $12$ appears to be a repetition of section $3$.

\subsection{Global statistics}
\begin{figure}[!ht]
\centering     \includegraphics[width=0.85\textwidth]{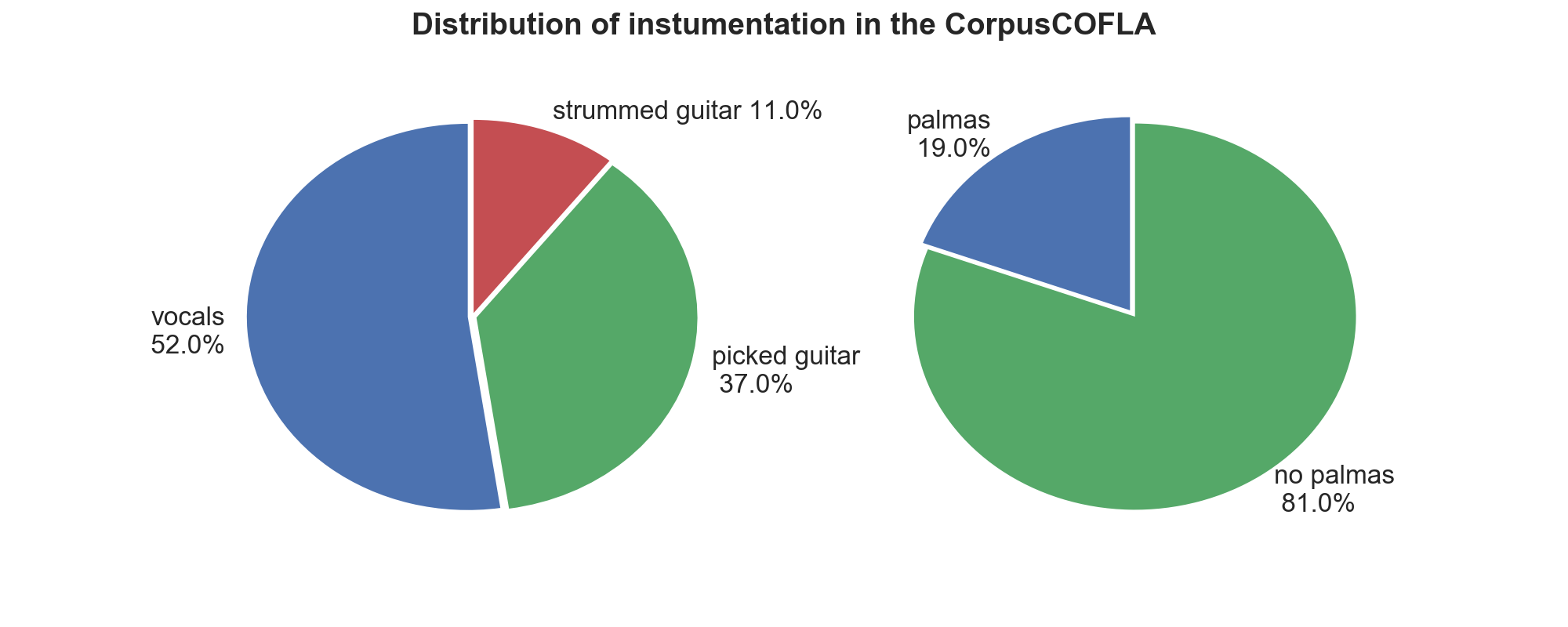}
      \caption{Global corpus statistics.}
       \label{fig:pie}
\end{figure}

The extraction of the automatic annotations for all the recordings of the collection enables the computation of global statistics related to the presence of  different instrumentation scenarios. Figure \ref{fig:pie} shows  the  distribution of vocals, \textit{palmas} and picked and strummed guitar, computed over all recordings of the collection.
It can be seen, that only $52\%$ of the analyzed frames are estimated to contain vocals. This is an interesting observation, because  flamenco is known to be heavily centered around the singing voice, the core element and origin of the flamenco tradition. However, at the same time, this is a meaningful result, because many performances contain long, solo guitar introductions and interludes. The remaining $48\%$ of solo guitar sections can be further divided to $11\%$ of strummed and $37\%$ of picked guitar. This observation can be explained by the fact that the reported distribution refers to solo guitar sections only, where guitarists are given the opportunity to perform the so-called \textit{falsetas}, i.e., long melodic guitar introductions or interludes with an autonomous compositional identity. \textit{Falsetas} (\cite{falseta1, falseta2}) do usually contain dominant melodic lines, and are mainly picked, rather than strummed. Furthermore, \textit{palmas} are detected in $19\%$ of all frames. Even though there are no strict rules about the presence of \textit{palmas}, they are unlikely to be found in some of the styles played in slow tempo, including \textit{fandangos}, \textit{soleares}, \textit{siguiriyas} and \textit{malague{\~n}as}, which comprise a large part of the corpus.

\subsection{Discovery of instrumental and a cappella recordings}
Although a flamenco music performance usually contains both singing voice and guitar accompaniment, certain styles, which belong to the \textit{ton{\'a}s} family, are traditionally performed a cappella. These styles are of particular interest to musicological studies, since they are assumed to represent the origin from which flamenco evolved to its present form. Consequently, a number of musicological and computational studies have targeted their particular characteristics \cite{moracomparative, thompson1985flamenco, kroher2014computational}. However, assembling a representative corpus of a cappella flamenco recordings is not a trivial task, because the editorial metadata of commercial recordings do not contain information on the existence or absence of guitar accompaniment and style annotations are most often missing  in online repositories and audio sharing platforms. As a support to this claim, so far, there only exists one such dataset of $72$  manually assembled recordings of the  \textit{ton{\'a}s} family\footnote{\url{https://www.upf.edu/web/mtg/tonas}}.

Here, we aim to  discover automatically a cappella recordings within the \textit{corpusCOFLA} collection. To this end, we begin by assuming  that, due to the absence of the guitar, the frames of an a cappella recording will be mostly classified to the vocal class, taking into account a $5\%$ of classification error. Therefore, to retrieve  the a cappella recordings, we first set a high threshold  for the percentage of vocal frames that are detected in a recording and we evaluate the results. Then we lower the threshold by $1\%$  each time and we repeat the retrieval operation. Each time, we evaluate the results by reporting the retrieval precision, which indicates the fraction of relevant instances among all retrieved recordings. We also note the total number of retrieved songs.

\begin{figure}[!ht]
\centering
     \includegraphics[width=0.95\textwidth]{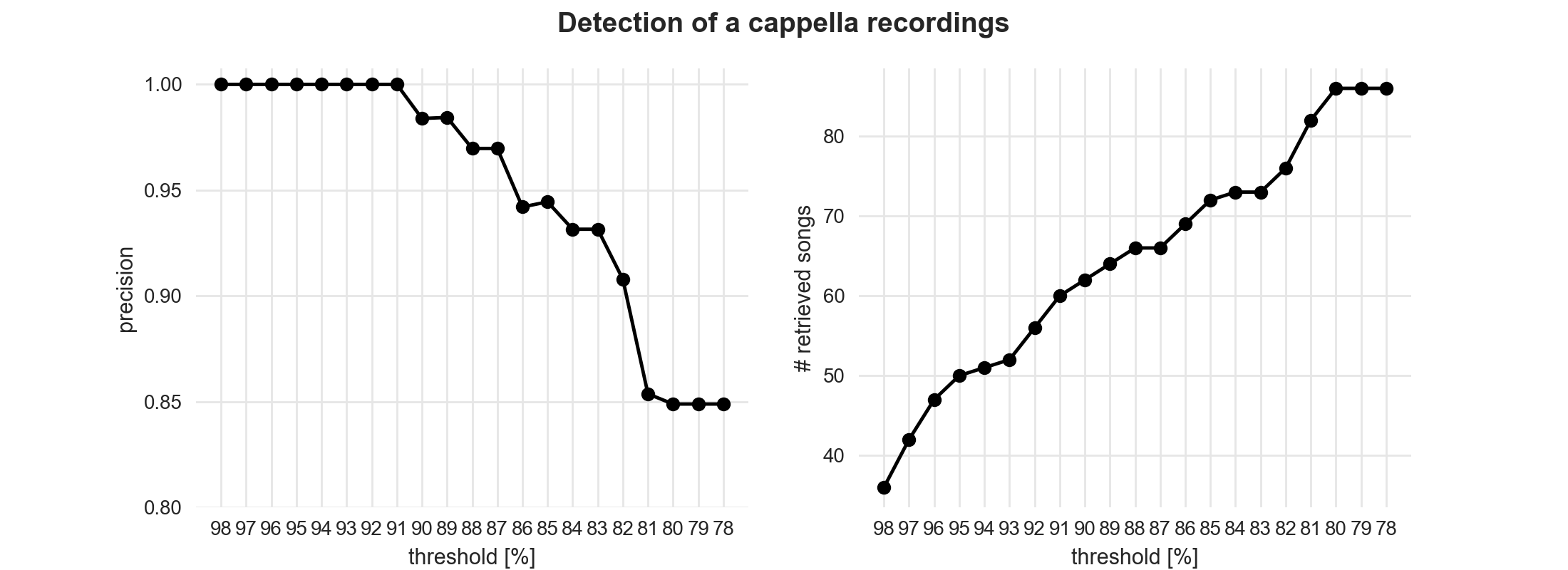}
      \caption{Discovery of a cappella recordings: Retrieval precision and total number of retrieved songs for varying thresholds.}
       \label{fig:acappella}
\end{figure}

The results in Figure \ref{fig:acappella} show that, when the threshold value is at least $91\%$, none of the retrieved recordings contain instrumental accompaniment and are therefore true hits. In this way, we identify $60$ a cappella recordings. When the threshold is further decreased, some more a cappella recordings are retrieved to the expense of an increasing number of false positives. A manual inspection showed, that the false positives are performances which  contain a guitar background in the presence of vocals, i.e.,  not in solo guitar interludes, hence the high percentage of vocal frames. This is also due to the setup of our vocal classifier, according to which, a frame is classified in the vocal class as long as it contains vocals, i.e, even when there exists simultaneous guitar accompaniment. For a threshold value of $78\%$, we retrieve a total of $86$ recordings, out of which $73$ are a cappella. 

\begin{figure}[!ht]
\centering     \includegraphics[width=0.95\textwidth]{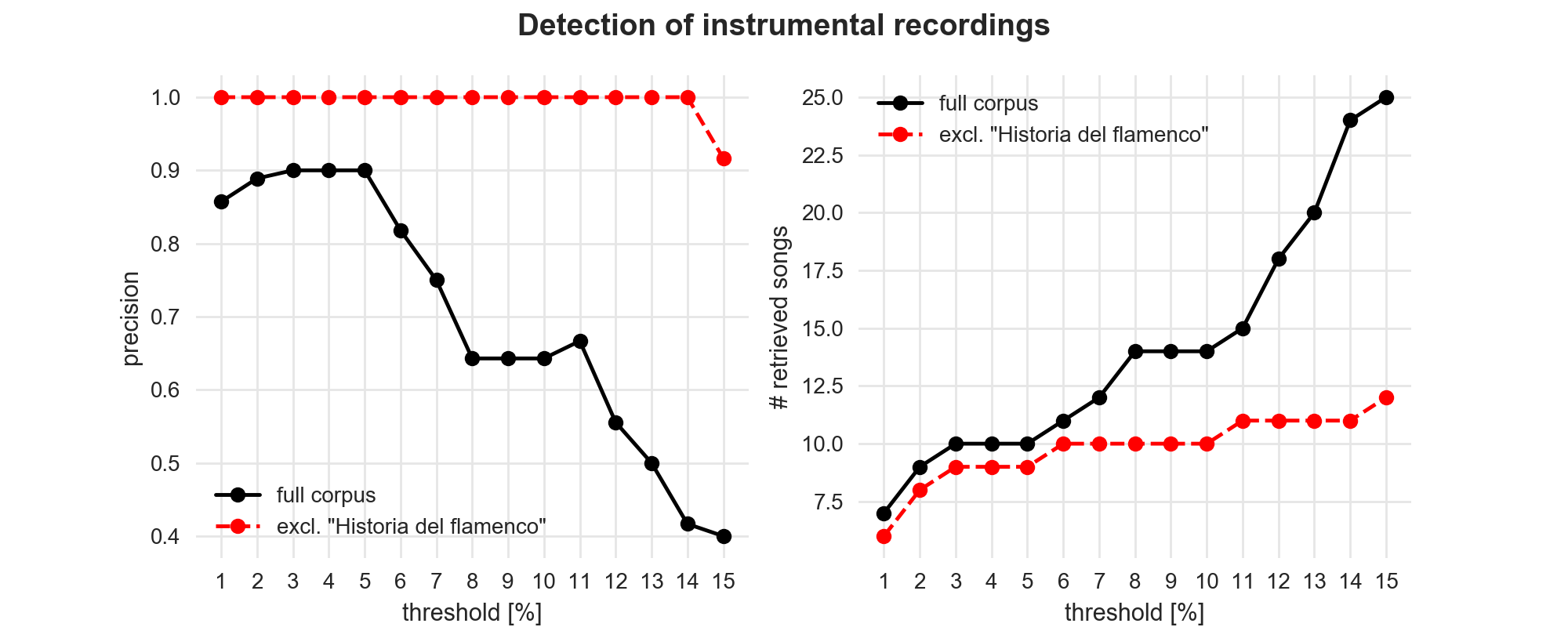}
      \caption{Discovery of instrumental recordings: Retrieval precision and total number of retrieved songs for varying thresholds.}
       \label{fig:instr}
\end{figure}

Similarly, we approach the discovery of instrumental recordings by analyzing songs with a low percentage of vocal frames. As in the previous experiment, we compute the retrieval precision for varying threshold values. In the case of instrumental recordings, the threshold refers to the maximum allowable percentage of vocal frames, i.e., the percentage below which a recording is considered to be instrumental. The results in Figure \ref{fig:instr} (black curves) show that for threshold values below $6\%$, around $90\%$ of the retrieved recordings are instrumental. However, the precision drops rapidly when the threshold increases further. A closer inspection revealed, that the large majority of false positives originate from the anthology \textit{Historia del flamenco}. This collection contains mainly historic field recordings of poor quality. In some of these tracks, the classifier failed to detect most of the singing voice sections, yielding a low percentage of vocal frames. Based on this observation, we repeated the experiment, excluding this particular anthology from the search space. In this new scenario (Figure \ref{fig:instr}, red curve), all retrieved recordings are instrumental performances for threshold values up to $14\%$. In this way, we were able to identify $11$ instrumental recordings in the corpus. As mentioned in Section \ref{sec:intro}, instrumental performances are rare in classical flamenco and it can be stated that solo guitar pieces have only recently gained popularity \cite{flamencoNuevo}.

These two experiments have shown that the structural annotations can be used for the semi-automatic (if perfect precision is required) and automatic (if a percentage of false alarms can be tolerated) identification of a cappella and instrumental recordings in large flamenco collections. The resulting metadata represent an important augmentation to the commonly provided editorial meta-data scheme, which usually lacks this type of information. Furthermore, the creation of collections of a cappella recordings with minimal manual intervention is of great importance for scaling existing computational approaches. Furthermore, the automatic tagging of instrumental recordings is important for genre-specific recommender systems, because solo guitar flamenco recordings may appeal to a different audience than classical flamenco, which mainly evolves around the singing voice.

\subsection{Differences in structure and instrumentation across styles}
Flamenco styles are distinguishable by a complex set of harmonic, melodic and rhythmic characteristics. Computational approaches to automatic style classification have so far been limited to a small number of substyles which can be identified based on common melodic templates \cite{styleClass1,styleClass2}. Consequently, the problem of automatically detecting the style of a flamenco recording remains largely unresolved \cite{styleDetection}. 

The occurrence of different instrumentations and the existence of prototypical structures across styles have not yet been formally studied. Therefore, in order to gain a better insight into style-specific characteristics and evaluate the potential use of the proposed structural segmentation backend for the purposes of automatic style classification, we investigate if (a) certain types of instrumentation are more likely to occur in particular styles and (b) songs belonging to the same style are similar in structure.

To this end, each song is represented as a feature vector, where each feature dimension stands for the percentage of vocals, \textit{palmas} and strummed and picked guitar. This representation is used to assess differences across styles on a multi-dimensional plane. In addition, we investigate instrumentation as a temporal structure, to explore intra- and inter-style structural similarity. Specifically, the evolution of instrumentation throughout a recording is represented by a sequence of feature vectors, where each vector encodes the classification decisions of the backend at the short-term level. Both representations are applied on the ten most frequently occurring styles of the corpus (Table \ref{tab:top10styles}) and recordings with multiple styles are excluded.

\subsubsection{Pair-wise visualization of song-level descriptors}
\begin{figure}[!ht]
\centering
     \includegraphics[width=0.99\textwidth]{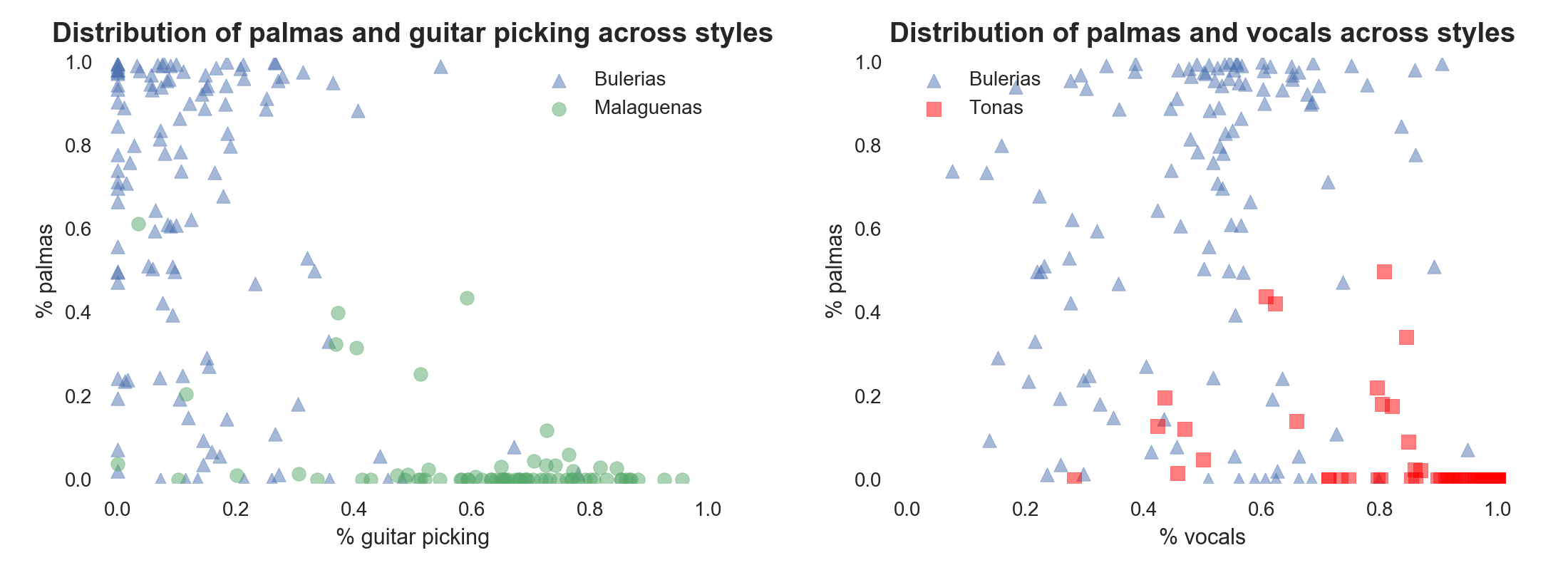}
      \caption{left: \% of picked guitar frames (x-axis) and \% of \textit{palmas} frames (y-axis) for \textit{buler{\'i}as} and \textit{malague{\~n}as}; right: \% of vocal frames (x-axis) and \% of \textit{palmas} frames (y-axis) for \textit{buler{\'i}as} and \textit{ton{\'a}s}.}
       \label{fig:styleScatter}
\end{figure}

In a first experiment, we visualize pairs of styles with respect to the song-level percentages of \textit{palmas}, vocals and picked guitar, and, for some cases, we observe that pairs of these features exhibit style-specific behavior. Two examples are shown in Figure \ref{fig:styleScatter}. The left scatter plot shows the percentage of picked guitar vs. the percentage of \textit{palmas} for the styles of \textit{buler{\'i}as} and \textit{malague{\~n}as}. It can be seen that the two styles are almost linearly separable on the respective two-dimensional plane. While most  \textit{buler{\'i}as} exhibit a high percentage of \textit{palmas}, the \textit{palmas} are largely absent in the \textit{malague{\~n}as} style. Regarding the guitar playing technique, picking is dominant in \textit{malague{\~n}as}, whereas the \textit{buler{\'i}as} mainly contain strummed guitar. 

Using our prior knowledge on a cappella styles, we now compare recordings of \textit{buler{\'i}as} to the \textit{tonas} family, which encompasses most a cappella styles of flamenco music. Figure \ref{fig:styleScatter} (right) presents the joint distribution of the percentage of vocals and \textit{palmas} for the two classes. As expected, the a cappella recordings contain a high percentage of vocal frames, mostly above $90\%$ and do not contain \textit{palmas}. Some outliers can be explained by the fact that one of the a cappella styles, the \textit{martinete}, is usually accompanied by stokes of a hammer on a anvil. A manual inspection revealed that this accompaniment was often mistakenly classified as either \textit{palmas} or picked guitar (due to the harmonic nature of the produced sounds). The combination of the low percentage of vocals and the presence of \textit{palmas} in the \textit{buler{\'i}as} leads again to a practically linearly separable  visualization of the two styles.

\subsubsection{Two-dimensional distance matrix visualization}
\label{subsubsec:single_vector}
Based on these observations, we now aim to understand how well this feature representation of a single vector per recording can discriminate the ten most frequent styles in the corpus. Therefore, we compute pair-wise Euclidean distances among all involved recordings and visualize the two-dimensional \textit{ForceAtlas2}-layout \cite{forceAtlas} of the resulting distance matrix. The \textit{ForceAtlas2} algorithm models distances as forces, causing repulsion between instances. Consequently, the resulting layout reflects the similarity between instances via the proximity in the designed two-dimensional space. 

\begin{figure}[!ht]
\centering
     \includegraphics[width=0.7\textwidth]{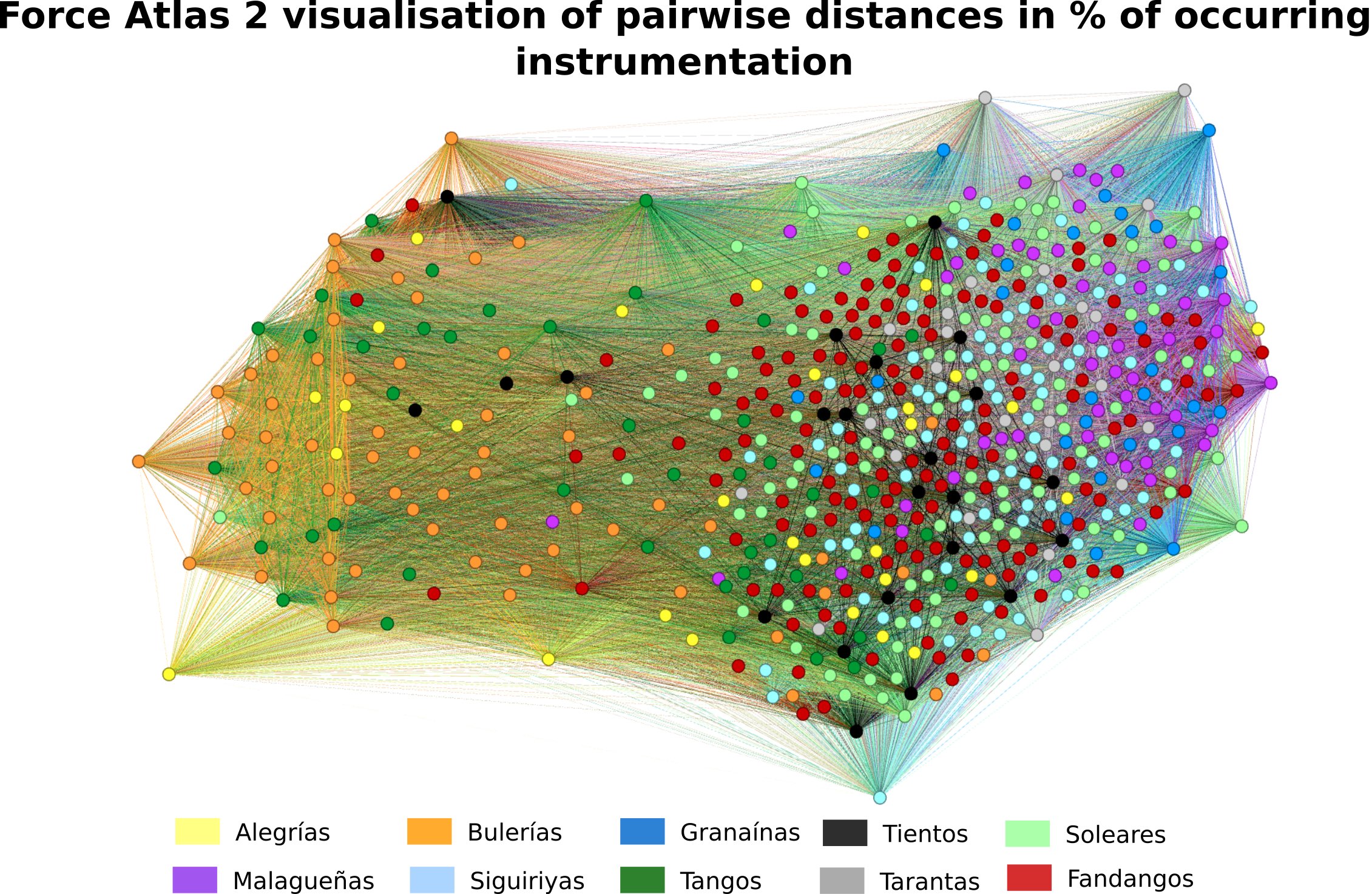}
      \caption{Force Atlas 2 layout of pairwise distances between song-level descriptors.}
       \label{fig:gephiStats}
\end{figure}

The resulting layout is shown in Figure \ref{fig:gephiStats}. It can be seen that two clusters are formed, one of which contains mainly recordings belonging to the styles of \textit{buler{\'i}as} and \textit{tangos}. The remaining styles are located in a second cluster. Within this group, we observe a slight tendency of the \textit{malague{\~n}as} being clustered together and the \textit{fandangos} being closer to the second cluster. The \textit{alger{\'i}as} appear to spread over both groups. 

\subsubsection{Structural similarity across styles}
\label{subsubsec:DTW}
At a next step, we investigate intra-style and inter-style structural similarity, i.e., we extend the previous experiment by taking into account the aforementioned temporal representation of the extracted annotations. Each recording is now represented as a $3$-dimensional sequence, where each dimension corresponds to the frame-wise binary decisions of the respective CNN. For each pair of sequences, a \textit{dynamic time warping} (DTW) \cite{dtw} cost is computed. To avoid excessive time stretching during the DTW operation, the global path constraint in \cite{itakura} is employed. Consequently, the distance between two sequences is equal to the warping cost of their alignment, that may be infinite due to the adopted global constraint. In order to reduce the computational burden of the DTW computations, each sequence is sub-sampled prior to the application of the alignment operation. In particular, one $3$-d vector is preserved every $1$ second. Finally, as it was the case with the previous experiment, the \textit{ForceAtlas2} layout is generated. 

\begin{figure}[!ht]
\centering
  \includegraphics[width=0.95\textwidth]{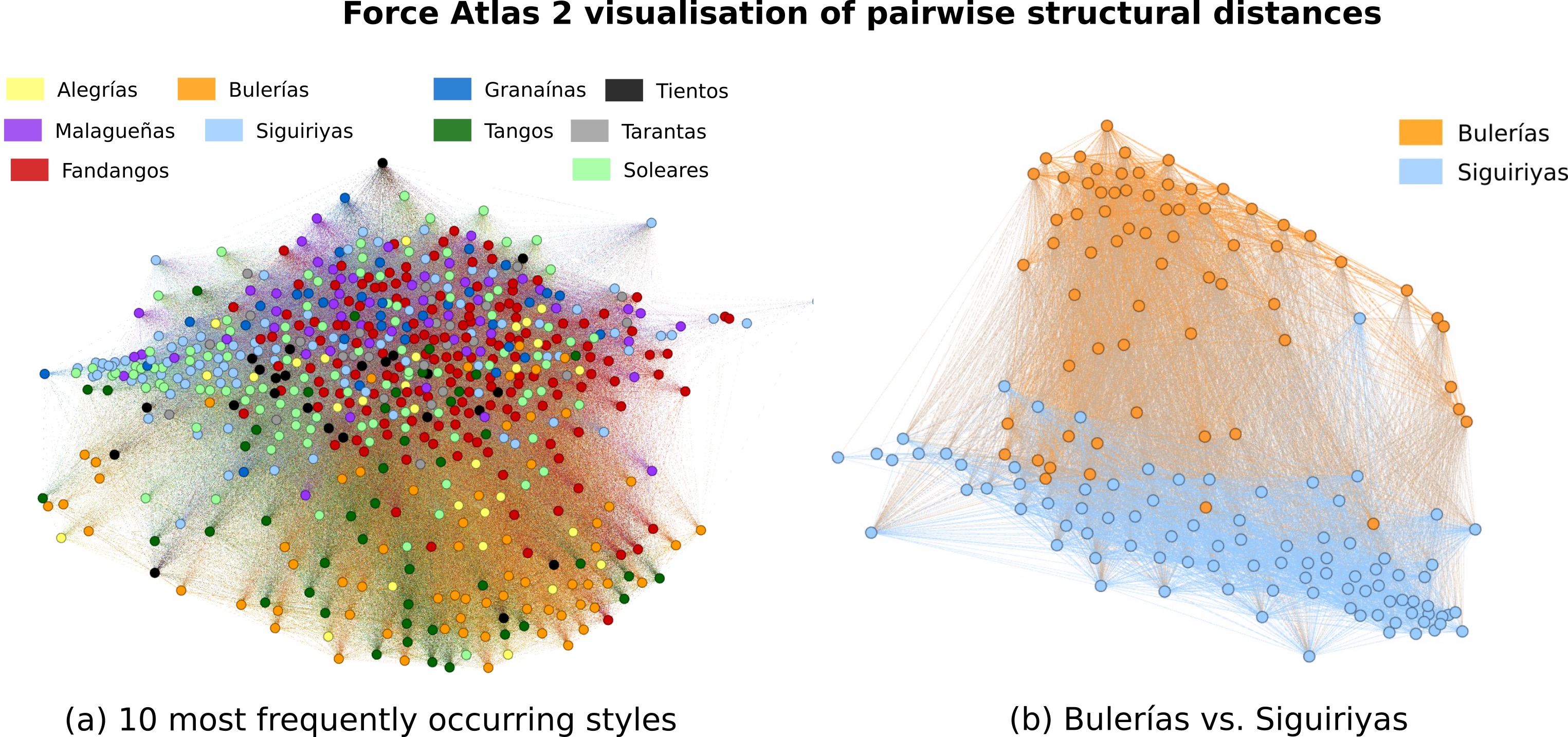}
      \caption{Force Atlas 2 layout of pairwise structural distances: (a) 10 most frequently occurring styles; (b) \textit{buler{\'i}as} and \textit{siguiriyas}.}
       \label{fig:gephi1}
\end{figure}

The resulting visualization is now shown in Figure \ref{fig:gephi1} (a). The structure of the layout is similar to the results of the previous experiment (Figure \ref{fig:gephiStats}). Again, \textit{buler{\'i}as} and \textit{tangos} appear separately from the other styles. However, the internal structure of the second, large cluster, differs slightly from the previous experiment. In particular, the \textit{soleares} and \textit{seguiriyas}, which were previously spread across the cluster, now tend to group together, indicating temporal structural similarity. 

Even though the experiments of Sections \ref{subsubsec:single_vector} and \ref{subsubsec:DTW} show, that the (dis)similarity of instrumentation-related representations is not sufficient to discriminate among all ten investigated styles, they do however demonstrate the potential of distinguishing between pairs of styles, as it was shown in Figure \ref{fig:styleScatter}. This is also supported by the example of Figure \ref{fig:gephi1} (b), where \textit{buler{\'i}as} and \textit{siguiriyas} form clearly separated clusters, when the DTW cost is computed for all pairs of recordings of these two styles.

\subsubsection{Distance-based style retrieval}
\label{subsubsec:MRR}
 \begin{figure}[!ht]
\centering   \includegraphics[width=0.95\textwidth]{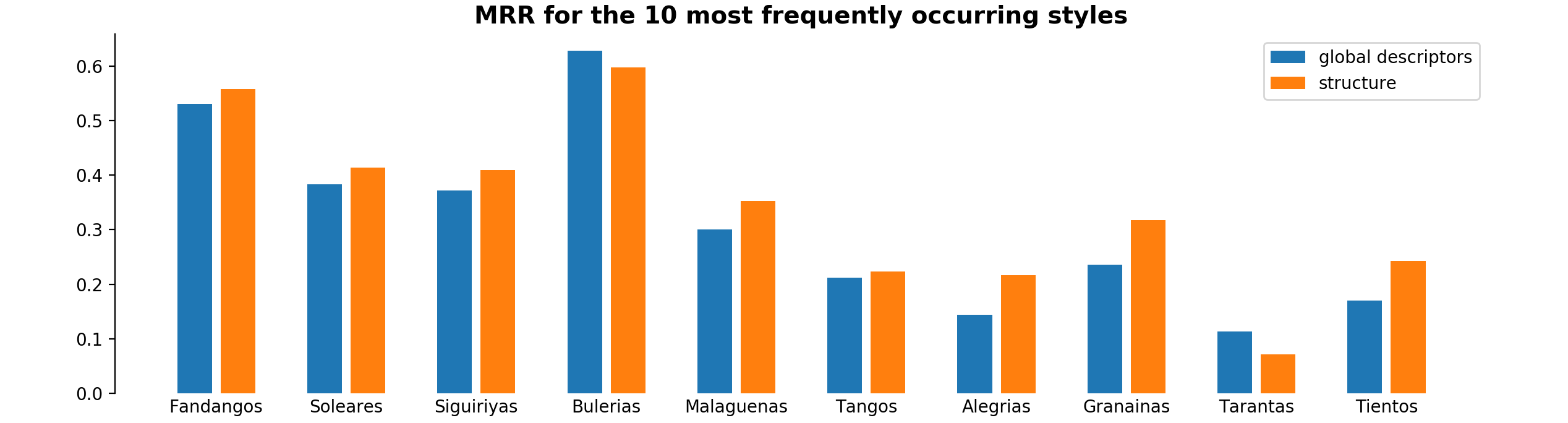}
      \caption{MRR for the distance-based retrieval of the ten most frequent styles, when using song-level descriptors vs. structural annotations.}
       \label{fig:mrr}
\end{figure}

 In order to enhance our understanding of the discriminative capabilities of the computed distance values, we now formulate a distance-based retrieval task. Namely, for each recording, we retrieve the ten most similar recordings and analyze their style with respect to the style of the query.  The success of the retrieved results is quantified, for each style, via its \textit{mean reciprocal rank}, $MRR_{style}$, which is defined as follows:

\begin{equation}
\mathrm{MRR}_{\mathrm{style}} =  \frac{1}{M_{\mathrm{style}}}\sum_{m=1}^{M_{\mathrm{style}}}\frac{1}{\mathrm{rank}_m}
\end{equation}

\noindent where $M_{\mathrm{style}}$ denotes the number of recordings belonging to the style and $\mathrm{rank}_m$ is the highest position in the list of returned results of the result that matches the style of the query, when the $m^{th}$ recording of the style is used as a query, assuming that the top result is found in position $1$. The MRR values have been computed for both types of representations, i.e., the single-vector representation per recording and the feature sequence one. The results in Figure \ref{fig:mrr} indicate that the DTW cost yields, for most styles, a higher MRR. Interestingly, in both cases, the highest MRR is observed for the style of \textit{buler{\'i}as}. This result is in line with the observations drawn in previous sections, where it was shown that this particular style exhibits certain characteristics that distinguish it from most other styles. It should also be noted that the reported results are influenced by the imbalance of styles in the corpus. For example, the \textit{fandangos} style accounts for nearly half of the search space. This explains its relatively high MRR, despite the rather low separability of this style, as it was shown in the previous sections via the \textit{ForceAtlas2} approach.

In conclusion, this exploratory investigation has shown, that song-level and local descriptors related to instrumentation hold valuable information with respect to style identity. Even though these features alone are not sufficient to distinguish between the styles investigated in this study, they do however show potential for their application, together with features related to other musical dimensions, in automatic style detection systems. Furthermore, they allow us to formalize typical characteristics of particular styles in a data-driven manner. 

\subsection{Tonality across instrumentation and styles}
As a last experiment, we illustrate flamenco-specific music theoretical concepts about the relationship between style, instrumentation and tonality in a large-scale data analysis environment. 
In flamenco, we encounter, apart from major and minor mode, a third mode, often referred to as the \textit{flamenco mode}. Most styles are restricted to on of these three modes. For example, alegr{\'i}as, are set in major mode, and \textit{seguiriyas} are set in \textit{flamenco mode}. The \textit{fandangos} are an exception, because they are described in the literature as bimodal, in the sense that guitar solo sections are performed in \textit{flamenco mode}, while singing voice sections are performed in major \cite{bimodalidad}. However, previous computational studies \cite{styleDetection} on a small dataset have indicated a weaker major mode identity during singing voice sections for this style, compared to other styles. Here, we study this phenomenon in a data-driven approach on a larger scale. Specifically, we analyze the estimated tonality for three styles, i.e., the \textit{fandangos}, \textit{seguiriyas} and \textit{alegr{\'i}as}, in vocal, strummed and picked guitar sections. 

A common method to determine the tonality of a piece is to extract its \textit{pitch class profile} and compare it to tonality-specific \textit{pitch class templates} \cite{template}. Pitch class profiles quantify the  occurrence of scale degrees throughout a composition. The statistical occurrence of scale degrees is closely related to the perceived tonality \cite{tonality} and pitch class templates are prototypical scale degree occurrence patterns, specific to a particular tonality. These are either extracted from large annotated corpora or are estimated through listening experiments \cite{tonalityExperimental, template}. If the pitch class profile of a given piece exhibits a high correlation with the pitch class template, it is likely that the piece will be perceived as belonging to the tonality associated with the template. 

In line with this approach, we estimate the tonality of a flamenco recording by correlating its pitch class profile with tonality-specific templates. To this end, we first extract \textit{harmonic pitch class profiles} (HPCP) \cite{hpcp} on a frame-level basis (non-overlapping frames of length $4096$ samples), and then average over vocal, strummed and picked guitar segments. In this way, we create three pitch class profiles for each recording. 

For the major mode, we use the template from \cite{template}. Furthermore, we extract a flamenco mode template from all recordings in the corpus belonging to the \textit{soleares} style. This process requires key normalization per recording,  in order to account for possible transposition among performances. Therefore, we normalize each pitch class histogram to a randomly selected reference example, by computing the correlation between circularly shifted versions of the histogram and the reference. For each recording, the pitch shift yielding the highest correlation is applied to all HPCP vectors.

\begin{figure}[!ht]
\centering
\includegraphics[width=0.65\textwidth]{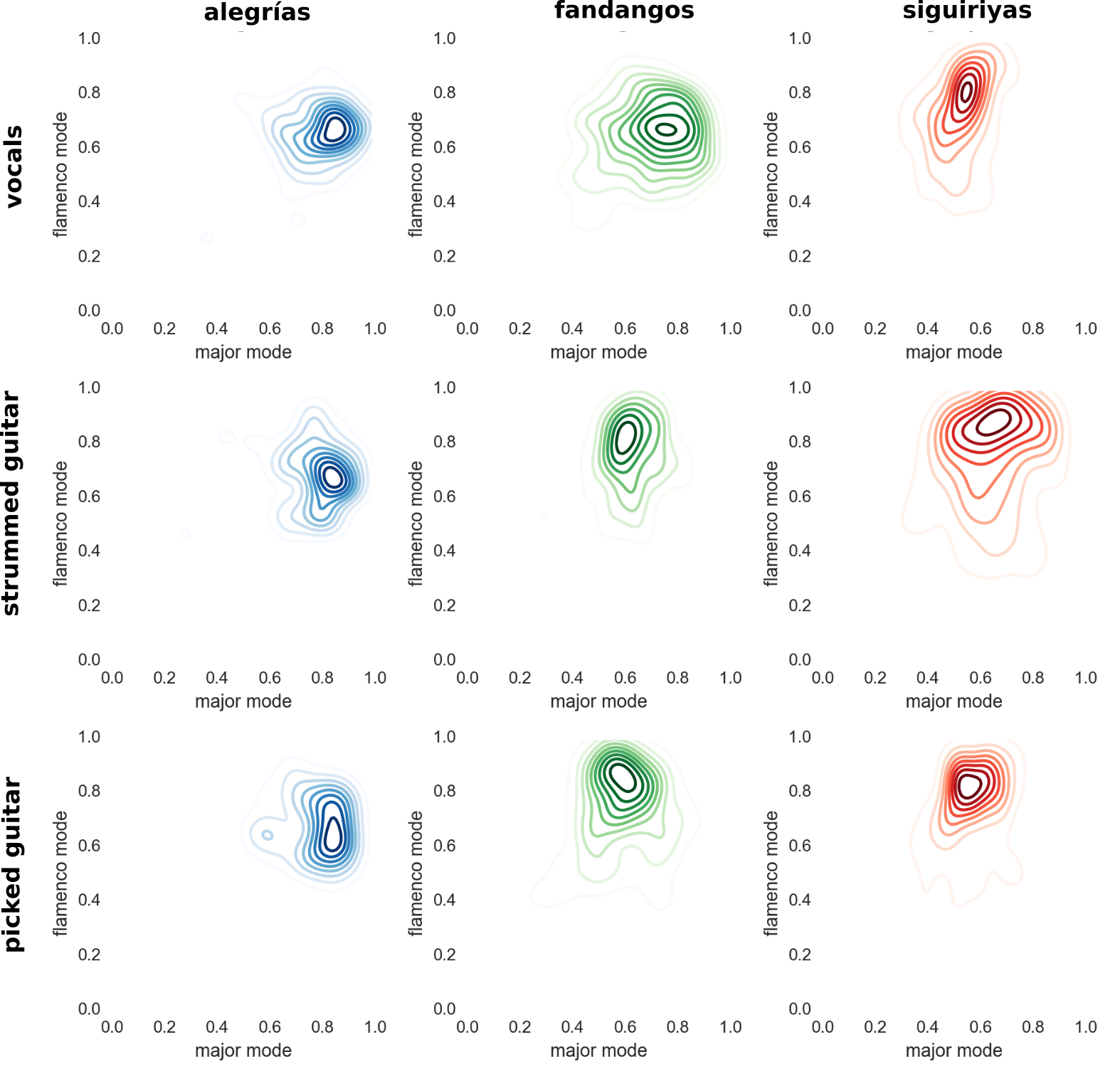}
      \caption{Kernel density plots displaying the correlation of pitch class profiles with pitch class templates across styles and instrumentation.}
       \label{fig:tonality}
\end{figure}

We then compute the correlation between the pitch class profiles of the recordings under study and both templates. Here, we again normalize each profile to the key of the respective template. The results, broken down into styles and instrumentation, are shown in the form of kernel density plots in Figure \ref{fig:tonality}. High values on the x-axis correspond to a high correlation with the major mode template and consequently indicate a strong major mode identity. Similarly, high y-values correspond to a strong flamenco mode identity. 

It can be seen, that across instrumentation, the \textit{alger{\'i}as} show a higher correlation with the major mode template, and the \textit{siguiriyas} with the flamenco mode template. This observation is in line with  theoretical studies \cite{fernandez}. For the style of \textit{siguiriyas}, we  observe less variance for picked guitar sections, compared to vocal and strummed guitar sections. In the plots displaying the \textit{fandango} recordings, we can see the bimodality of the style. Specifically, both  picked and strummed guitar sections exhibit a higher correlation with the flamenco mode template. In contrast, the vocal sections yield a higher correlation with the major mode template. However, similar to the findings in \cite{styleDetection}, we observe a weaker correlation and a larger variance, compared to the \textit{alegr{\'i}as}. 

This experiment demonstrates the theoretically formalized bimodality of the \textit{fandango} style, and, at the same time, reveals two interesting aspects related to tonality in flamenco. First, we observe a stronger tonal identity in picked compared to strummed guitar sections in \textit{seguiriyas}. Secondly, we confirm the observations from \cite{styleDetection} on a larger scale, which indicate that the vocal sections of the \textit{fandangos} exhibit a weaker major mode identity compared to the \textit{alegr{\'i}as}.

\section{Conclusion}\label{sec:conclusions}
The first goal of this paper was to develop state-of-the-art classifiers that are capable of providing reliable, content-based instrumentation-related annotations of flamenco recordings. The second goal was to use the extracted annotations to enable the realization of musicological findings that would otherwise require time-consuming manual procedures, and verify, at a large scale, via computational means and a data-driven approach, existing musicological observations. The paper showed that the developed annotation backend is reliable enough to achieve the goals that were set and this was demonstrated via a number of related experiments that investigated the discriminative capability among genres of higher-level features drawn from the extracted low-level descriptors of instrumentation. The reported results focused both on pairwise style relationships and multi-style associations, via a variety of complementary approaches, ranging from visualizations of varying complexity to a style-based retrieval engine. In all experiments, emphasis was given on explaining the findings from a computational and musicological perspective. Our computational study is the first of its kind in the context of flamenco music. The developed backend is provided in the form of pre-trained models for the sake of reproducibility of results and as a tool for studies in related ethnomusicological tasks.

\section*{Acknowledgements}
\begin{itemize}
\item We gratefully acknowledge the support of NVIDIA Corporation with the donation of the Titan Xp GPU used for this research.
\item This research has been partly funded by the Junta de Andaluc{\'i}a and the FEDER funds of the European Union, project COFLAII under the grant identifier P12-TIC-1362 and the international mobility grant of the University of Seville (plan propio de investigaci{\'o}n y transferencia 2017-1.3-A). 
\end{itemize}

% Bibliography
\bibliographystyle{alpha}
\bibliography{mining}
\end{document}